\documentstyle{mn}
\newcommand{\othree}{[O{\sc iii}]\,}
\newcommand{\fetwo}{[Fe{\sc ii}]\,}
\newcommand{\otwo}{[O{\sc ii}]\,}
\newcommand{\oone}{[O{\sc i}]\,}
\newcommand{\hone}{H{\sc i}\,}
\newcommand{\ntwo}{[N{\sc ii}]\,}
\input{psfig}

\title[Observations of the neutral gas and dust in the radio galaxy 3C\,305]
{Observations of the neutral gas and dust in the radio galaxy 3C\,305 }

\author[Jackson et al.]
{N. Jackson$^{1}$, R. Beswick$^1$, A. Pedlar$^{1,3}$,G.H. Cole$^1$, 
W.B. Sparks$^{2}$, J.P. Leahy$^{1}$,\cr D.J. Axon$^{4}$,A.J. Holloway$^{1}$ \\
$^{1}$University of Manchester, Jodrell Bank Observatory,
Macclesfield, Cheshire SK11~9DL, UK \\ 
$^{2}$Space Telescope Science Institute, San Martin Drive, MD21218, USA\\
$^{3}$Onsala Space Observatory, Chalmers Institute, S-43992 Onsala,
Sweden\\
$^{4}$University of Hertfordshire, Department of Physical Sciences,
College Lane, Hatfield, Herts, AL10 9AB
} 

\date{06-05-2002}
\def\ltsim{\ifmmode\stackrel{<}{_{\sim}}\else$\stackrel{<}{_{\sim}}$\fi}
\def\gtsim{\ifmmode\stackrel{>}{_{\sim}}\else$\stackrel{>}{_{\sim}}$\fi}
\begin{document}
\input{epsf}
\maketitle

\begin{abstract}

We present MERLIN and Hubble Space Telescope (HST) 
observations of the central region of the nearby
radio galaxy 3C\,305 and use them to study the gas and dust in this
object. The MERLIN observations are of neutral hydrogen (\hone )
absorption against the strong non-thermal 20cm continuum seen towards the
central 4~kpc  of 3C\,305. Our $\sim$0\farcs2 (160 pc)
resolution observations show that the \hone\ absorption is highly localised
against the south-western radio-emission with column densities
$\sim$1.9$\times$10$^{21}$ cm$^{-2}$. The absorption is broad (full
width at half maximum, FWHM, of 
145$\pm$26 km s$^{-1}$) and red-shifted by 130 km s$^{-1}$ relative to the
systemic velocity. 
The HST images in multiple optical and
infrared filters (430\,nm, 702\,nm, \othree 500.7\,nm, \fetwo 1.64$\mu$m
and $K$-band polarization) are presented. Evidence is seen for coincidence
of the \fetwo emission with the knot at the end of the radio jet, which
is evidence for the presence of shocks.

We compare the optical and radio images in order to investigate the
relationship between the dust and neutral gas distributions. An 
unresolved (0\farcs07) nucleus is detected in $H$ and $K$ and its
properties are consistent with a quasar reddened by $A_{V}>4$. We
propose that the absorption arises in a region of neutral gas and dust.
Its structure is complex but is broadly consistent with an inclined disc
of gas and dust encircling, but not covering, the active galactic nucleus.
A comparison of the neutral gas observations and previous emission-line 
observations suggests that both the neutral and ionised gas are undergoing
galactic rotation towards the observer in the north-east and away from 
the observer in the south-west. We propose that the outflow giving rise
to the radio emission has a component towards the
observer in the north-east and away from the observer in the south-west.
Unfortunately as we do not detect radio emission from the compact nucleus
we cannot set limits to neutral hydrogen absorption from a
circumnuclear obscuring torus.

\end{abstract}

% uncomment this for final version
%\Large

\section{Introduction}

The formation and evolution of active galaxies is an interesting and
unsolved question. It is not yet obvious why a small fraction of galaxies
have a degree of central activity many times larger than that of normal
galaxies, or why a still smaller fraction produce radio jets. Nor is it
fully understood why active galaxies harbouring the most powerful radio
sources appear to be almost exclusively elliptical galaxies. Much 
work has been done investigating the possibility that active galaxies form
as a result of mergers (Toomre \& Toomre 1972, Hutchings et al. 1984,
Canalizo \& Stockton 2001); the evidence for this
includes disturbed morphologies, including features suggestive of tidal
tails. More recently, Hubble Space Telescope (HST) imaging of a large
sample of 3CR galaxies (de Koff et al. 2000) has revealed that a large
number of nearby radio-loud active galaxies contain prominent dust lanes,
suggestive of the capture of interstellar medium by the accreting galaxy. 

The study of the physical properties of the dust and gas in relativistic
plasmas at the centres of active galaxies is of great interest. The dust
distribution can be studied using optical and infra-red imaging. The
atomic gas can be studied using the 21-cm line of neutral hydrogen. As
well as giving clues to formation of active galaxies, dust and gas may
obscure the direct view to the active galactic nuclei (AGN) and thus
affect our ability to view the central optical continuum and
broad-emission-line region. Unfortunately,
emission lines from atomic hydrogen can only be seen by integrating over
large areas --- typically tens of arcseconds --- because of its low ($\sim
100$K) brightness temperature and the current sensitivities of decimeter
arrays. However, provided the neutral gas is in front of a strong source
of radio continuum, absorption studies can be used to study the atomic 
hydrogen on subarcsecond scales. Parameters such as the transverse 
velocity gradients and structure of the neutral gas can then contribute
to an understanding of the dynamics of the nuclear regions of
active galaxies.

3C\,305 is a low-redshift radio galaxy with several unusual properties.
The galaxy has some of the characteristics of a spiral galaxy with a thin
dust lane crossing the nuclear region and was classified by Sandage (1966)
as an Sa(pec). The radio structure is an order of magnitude smaller than
is usual for powerful, steep spectrum sources and reveals twin central radio
jets feeding two radio lobes, with extended weak emission from perpendicular
arms (Lonsdale \& Morison 1980, Heckman et al. 1982). The radio source is
confined within and perhaps by the galactic disc. Two investigations by
Heckman et al. (1982) and Jackson et al. (1995) have revealed a strong
association between the radio structure and the optical emission line
region. There is strong evidence (Jackson et al. 1995; Martel et al. 1999)
that much of the optical view of the source is affected by dust
obscuration. 

Heckman et al. (1982) suggested that the galaxy may be the result of a
merger which triggered nuclear activity. The remnants of the merger
debris are interacting with the radio jet to produce the striking radio 
morphology. There is also evidence for two tidal tails extending from the
galaxy (Heckman et al 1986), which may have led to Sandage's original
classification of the galaxy as Sa(pec), and for unusual stellar
kinematics. Heckman et al. (1982) found a low stellar velocity dispersion 
(170km\,s$^{-1}$), relatively high stellar rotation velocity (a mean 
140km\,s$^{-1}$ for 0.2-0.5$r_e$) about the minor axis (PA 7$^{\circ}$) 
which was of solid-body form to 5kpc and a flat rotation curve further 
out. They also found a ``rotation'' amplitude of over 200km\,s$^{-1}$ for
the \othree\ gas and concluded that the stellar velocity is only just over
50\% of the gas velocity. Finally, they identified the twisting of the 
optical isophotes with distance from the nucleus, and the presence of a dust
lane, as strong evidence for a merger origin of the galaxy. Peculiar 
kinematics fit naturally into this picture and would be expected if a 
companion is gradually being absorbed into the main body of the 3C\,305 
host galaxy. Furthermore, both ultra-luminous infrared galaxies (ULIRGs,
Sanders et al. 1988) and quasars (Heckman et al. 1986) are suspected to be
frequent consequences of mergers, with a possible evolutionary sequence 
(see e.g. Barnes \& Hernquist 1992) connecting mergers to ULIRGs and thence
to quasar activity on a timescale of a few hundred million years.
3C\,305, however, has a 60$\mu$m flux of 0.2~Jy, about 2\% of that of a
typical ULIRG at its redshift. Any merger must therefore have taken
place some time ago or else have produced relatively little of the
obscured star formation which accounts for the far-infrared flux of a
typical ULIRG.

More recently, Baum et al. (1988) have also observed the extended optical
emission-line gas and Draper et al. (1993) have studied the optical
polarization, which is probably dichroic in origin and associated with
the same foreground screen that obscures the optical view of the nucleus.
Jackson et al. (1995) used HST images of the emission-line gas to quantify
the radio-optical association and to constrain models of the interaction
of radio jets with the interstellar medium (ISM).

The relatively low redshift (z=0.0417) of 3C\,305 makes it an ideal
candidate to study the neutral gas at 0\farcs2-resolution using MERLIN,
and to study the dust and emission lines at approximately the same
resolution using the HST in the optical and infrared. In this paper we 
present new images from both instruments. The aims are to investigate
whether or not the active nucleus can be detected in the infrared and measure
the extinction in front of it; to study the interaction of the radio jet
and emission-line gas and the contribution of the former to the excitation
of the line emission; and to use the estimates of dust and gas columns
we derive to constrain the geometry of the system. Throughout this paper
we will assume H$_0$ = 75 km s$^{-1}$ Mpc$^{-1}$ which implies a distance
of 167 Mpc and hence 1 arcsecond is equivalent to 810 pc in 3C\,305.

\section{Observations}

\subsection{ MERLIN 21cm observations}

The 21-cm radio observations were made in June 1997, using the MERLIN
array, including the 76-m Lovell telescope, resulting in a maximum
baseline length of 217 km. Two 8-MHz bands, with opposite circular
polarizations were observed with 64 correlator channels per
polarization, resulting in a channel bandwidth of 125 kHz (26.3
km\,s$^{-1}$). The central frequency in channel 32 corresponds to a
heliocentric optical velocity of 12\,825 km\,s$^{-1}$ (z=0.0417).

3C\,305 was observed over a period of 14 hours, interspersed with
observations of a phase calibrator, 1442+637, and a bandpass
calibrator, 0552+398 was observed for 1 hour. The absolute flux
density scale was determined from an observation of 3C 286, which was
assumed to have a flux density of 15.0 Jy (Baars et
al. 1977). Initial editing and calibration was carried out using
standard MERLIN programs.

Further editing was undertaken and phase and bandpass calibration were
applied within {\sc AIPS}\footnote{Astronomical Image Processing System,
distributed by the U.S. National Radio Astronomy Observatory}. 
Individual telescopes were weighted according to
their gains and 1442+637 was found to be suitable for both phase and
bandpass calibration. The continuum image was formed by compressing
channels 3 to 61 followed by 4 passes of self-calibration. The \emph{uv}
data were Fourier transformed using uniform weighting and deconvolved from
the beam of the telescope using a {\sc CLEAN}-based algorithm (H\"{o}gbom 1974).

A naturally weighted spectral line cube was formed by subtracting 24
absorption free channels (channels 6 to 13, 17 to 28 and 58 to 61) from
the \emph{uv} data. The \emph{uv} data were then Fourier transformed and
deconvolved from the beam. The cleaned continuum was added back to the
continuum-subtracted cube and spectra at specific coordinates were
obtained.

%  Not used????
%A 2-cm observation of 3C\,305 was extracted from the VLA archive (P.I. van
%Breugel). This was the same image as used by Jackson et al. (1995) for
%comparison of the optical and radio data in that paper.

\subsection{ HST observations}

3C\,305 was observed with the HST's Near-Infrared Camera/Multi-Object
Spectrograph (NICMOS) with a variety of filters on 1998 July 16 and July
19. Wide-band images were obtained at 1.6$\mu$m with the F160W filter
(roughly corresponding to Cousins $H$-band), and at 2.2$\mu$m with the
POL0L, POL60L and POL120L filters; polarization information was therefore
obtained at the latter wavelength. Narrow-band images were obtained with
the F170M filter, which covers the redshifted infra-red \fetwo line
at 1.644$\mu$m, and continuum was subtracted from these images using
data taken with the F180M filter.

Other data, all observed after the COSTAR
installation which solved the HST spherical aberration problem, were
extracted from the HST archive. These data included optical observations
from the Wide Field \& Planetary Camera 2 (WFPC2), 
taken with the F702W filter on 1994 September 04,
and including both continuum and H$\alpha$/\ntwo\ emission. A Faint
Object Camera (FOC) image at 430\,nm from 1997 January 12 was also
extracted. Table 1 contains complete details of the observations
including exposure times; figure 1 shows the contaminating lines which
are present in each continuum filter.

\begin{figure}
\psfig{figure=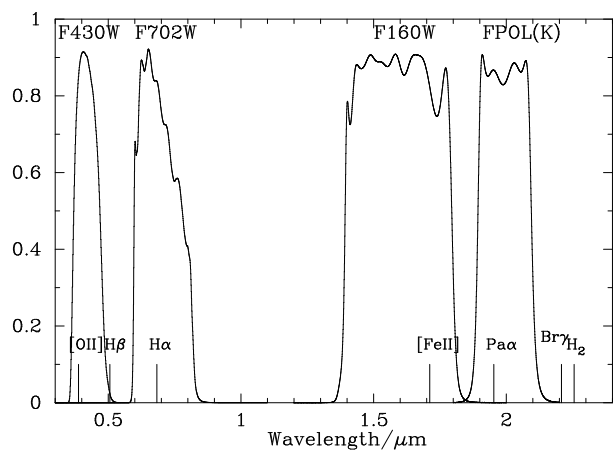,width=8cm,angle=-90}
\caption{Contaminating lines present in each of the HST continuum filter
bands used}
\end{figure}

\begin{table}
\begin{tabular}{lllrl}
Instrument & Filter & Date & Exp.  & Proposal \\
 & & & time /s & number\\
FOC    & F430W & 1997 Jan 12 & 297 & 6304*\\
WFPC2  & F533N (\othree )& 1995 Oct 13 & 600 & 5957*\\
WFPC2  & F702W & 1994 Sep 04 & 560 & 6254*\\
NICMOS & F160W & 1998 Jul 16 & 1088 & 7853\\
NICMOS & F172N (\fetwo )& 1998 Jul 16 & 3328 & 7853\\
NICMOS & F180N & 1998 Jul 16 & 3328 & 7853\\
NICMOS & POL0L 2$\mu$m & 1998 Jul 16 & 1088 & 7853\\
NICMOS & POL120L 2$\mu$m & 1998 Jul 16 & 1088 & 7853\\
NICMOS & POL240L 2$\mu$m & 1998 Jul 16 & 1088 & 7853\\
\end{tabular} 
\caption{Log of the HST observations. * denotes that the observation was
extracted from the HST archive.}
\end{table}

The reduction of optical data followed standard STScI pipeline procedures
including application of flatfields and distortion correction. The reduction of
the polarization images to Stokes Q and U images was 
performed using polarimetry routines described by Sparks \& Axon (1999).
Photometry was achieved
using the {\sc photflam} keywords in the file headers of the HST images. 

NICMOS images are affected by a ``pedestal'' or unpredictable bias,
which may be different in the four quadrants of the image. We used the
{\sc unpedestal} program of R. van der Marel to minimise this effect
(Bushouse, Dickinson \& van der Marel 2000). Before running {\sc
unpedestal} we temporarily masked out the bright galaxy cores, which
could affect the results, and fitted a single bias for each field or a
separate bias for each quadrant, depending on which showed the least
evidence for jumps in the background at the quadrant boundaries.
Accurate offsets between individual images, needed for making the final
mosaic, were obtained by fitting the galaxy peak; and the nominal
background was set to the mean of two standard regions well away from
the galaxy centre. Although galaxy light is present and hence will be
subtracted when the mosaic is constructed, the oversubtraction should be
identical in each image. These parameters were fed to the Space
Telescope Science Institute program {\sc calnicb} to
make the final mosaic; the background was re-determined and corrected
using regions at the edge of the mosaic. A small background gradient
remains after these corrections, but does not affect the results.

\subsection{Astrometry}

An archival HST WFPC2 F702W filter continuum image of 3C\,305 is shown in
Fig. 2.  The HST astrometry was tied to the International Coordinate
Reference Frame (ICRF) as follows. R. Argyle and
L. Morrison determined the position of four stars in the WFPC2 field by
re-measuring a 1981 Herstmonceux astrographic plate (Argyle \& Clements
1990), using the Royal Greenwich Observatory PDS machine; these positions
are listed in Table 2. Fourteen stars on the plate were observed by them
in 1997 with the Carlsberg Automatic Meridian Circle (CAMC); 
these measurements
are directly tied to HIPPARCOS positions, and hence allowed them to put
the WFPC2 stars onto the ICRF.

The standard header on the WFPC2 image was updated with the latest
calibration using the Space Telescope Data Analysis System task 
{\sc uchcoord}, and the four star positions
were measured with the task {\sc metric}, which accounts for the geometric
distortion of the image. A non-linear least-squares fit was made to find
the transformation required to align the two coordinate systems; we found
that the reduced $\chi^2$ was not significantly lowered by allowing for a
rotation and change of scale (best fit values for these differed by
0\farcs1 and 0.05\% from nominal), so we simply applied the best-fit
shift ($\sim$0\farcs4).  

The rms residual between astrographic and
corrected WFPC2 positions was 0\farcs12 in each coordinate, somewhat
larger than the expected uncertainties (the scatter may be dominated by
proper motion of the reference stars between 1981 and 1994). The final
registration should therefore be accurate to about 0\farcs1 and is shown
as an overlay in Fig. 2.

\begin{table}
\begin{tabular}{ccc}
Star & RA (J2000) & Dec (J2000)\\
1&  14 49 16.928$\pm$0.005  & +63 17 06.31$\pm$0.06\\
2&  14 49 14.985$\pm$0.006  & +63 15 54.34$\pm$0.05\\
3&  14 49 17.897$\pm$0.007  & +63 15 29.73$\pm$0.07\\
4&  14 49 28.873$\pm$0.017  & +63 15 39.43$\pm$0.07\\
\end{tabular}
\caption{CAMC star positions used to register the HST images with the ICRF}
\end{table}

\begin{figure}
\psfig{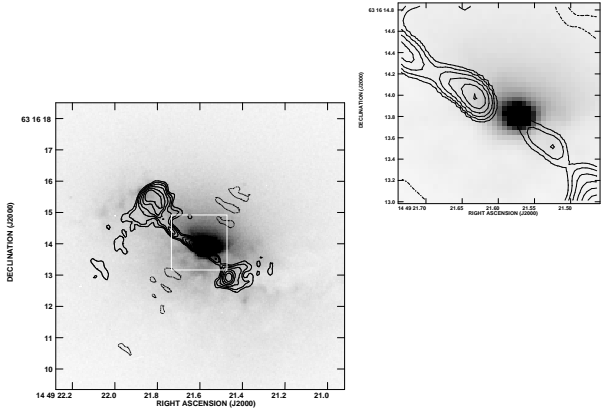}     
\caption{The radio map contours (see Fig. 3) overlaid on the F702W 
HST optical image (greyscale). The inset shows the area around the
centre of the galaxy. The relative astrometry is based on CAMC positions of 
four nearby stars as explained in the text and is uncertain to about 
0\farcs1. Contour levels are 1mJy/beam $\times$
($-$1,1,2,4,8,16,32).}
\end{figure}

Once an optical image had been aligned to the radio frame, the remaining
optical images were then aligned with each other using a star which
appeared about 12$^{\prime\prime}$ WNW of the optical centre and which
appears in all the optical HST frames. The
infra-red frames contained no suitable reference star. The IR images
were aligned with each other using the peak brightness. The IR-optical
alignment was performed by aligning the peak of the \fetwo emission
to the northeast of the core with the optical emission line peak. This also
aligned the $K$-band local maximum, corresponding to Pa$\alpha$ emission in
the northeast lobe, with the H$\alpha$ emission contaminating the optical
702-nm image. The resultant infra-red nuclear position is close to the
optical position (formally 63 mas E and 26 mas S of the optical
component, although this is close to the error [about 1 pixel, or 45   
mas] within which the images could be lined up). The 430-nm image also
did not contain the reference star, and so was lined up with the   
emission-line maximum in the northeast lobe. All images have been
rebinned to the Planetary Camera pixel scale of 0\farcs0455
for further analysis.

\section{Results}

\subsection{MERLIN observations of the 20-cm continuum emission}

%Improved a bit....

The 1.4-GHz continuum image of 3C\,305, derived from the line-free
channels, is shown in Fig. 3. The image has been derived from uniformly
weighted data and has an angular resolution of 0\farcs19$\times$0\farcs14
and a noise level of 0.10~mJy\,beam$^{-1}$. Note that the
extended emission, especially from the northeast and southwest 
arms, is affected by a
lack of MERLIN short-baseline spacings.The image reveals the structure of the
radio continuum emission with two jets forming radio lobes separated by
3\farcs6  in PA 54$^{\circ}$. Two low brightness arms extend
perpendicular to the orientation of the jets. In Table 3 we give a
summary of the 1.4-GHz flux densities of the continuum components.

The alignment of the radio and optical images (Fig. 2) is 
at first sight rather surprising, as it places the radio peak about
0\farcs45 away from the optical peak, nearly four times the astrometric 
error, in PA$\sim65^{\circ}$ from the brightest optical/infra-red point.
However, the central radio component does
not have a flat spectrum and is therefore unlikely to be the radio core;
it is likely that we are seeing a knot at the base of the jet,
with  a peak flux density of 13 mJy beam$^{-1}$. The
radio core is undetected and must therefore have a flux of $<$1\,mJy. 
On the other hand, the
astrometry places the northern termination of the radio jet almost
directly on top of a region of strong \othree\ emission, which is a
physically plausible result. Without the CAMC data, Jackson et al.
(1995) aligned the radio and optical maxima directly, resulting in the
head of the radio jet falling about 0\farcs5 short of the optical
emission line regions.

%\begin{figure*}
%\centering
%\psfig{figure=LCONT1,width=17cm,angle=-90}
%\caption{Contour map of the 1.4 GHz radio continuum emission from 
%3C\,305. 
%The map has a resolution of 0.19 $\times$ 0.14 arcsec and the contour
%levels 
%are 0.3$\times$(-1,1,2,4,8,16,32,64,128,256) mJy beam$^{-1}$.}
%\label{fig1}
%\end{figure*}

\begin{table*}
\centering
\caption{The 1.4 GHz continuum properties of the central region of 3C\,305.}
\label{tab1}
\begin{tabular}{lccccc}
\hline
\hline
Component    & S$_{1.4 \rm{GHz}}$ & Area         & Peak            & R.A. (J2000)               & Dec. (J2000)              \\
             & (mJy)              & (arcsec$^2$) & (mJy beam$^{-1}$)  & 14$^{\rm{h}}$49$^{\rm{m}}$ & 63$^{\circ}$16$^{\prime}$ \\
\hline
NE Arm       & 99.7               & 3.197        & 6.4             & 21$^{\rm{s}}$.860          &  14$^{\prime\prime}$.93  \\
NE Lobe      & 601.9              & 1.338        & 114.0           & 21$^{\rm{s}}$.832          &  15$^{\prime\prime}$.56  \\
NE Jet B     & 38.4               & 0.295        & 13.0            & 21$^{\rm{s}}$.753          &  14$^{\prime\prime}$.61  \\
NE Jet A     & 52.7               & 0.306        & 36.4            & 21$^{\rm{s}}$.634          &  13$^{\prime\prime}$.97  \\
SW Jet       & 9.4                & 0.288        & 4.4             & 21$^{\rm{s}}$.528          &  13$^{\prime\prime}$.52  \\
SW Lobe      & 188.0              & 0.860        & 80.2            & 21$^{\rm{s}}$.462          &  12$^{\prime\prime}$.93  \\
SW Arm       & 31.7               & 1.369        & 4.1             & 21$^{\rm{s}}$.396          &  12$^{\prime\prime}$.79  \\
             &                    &              &                 &                            &                           \\
\hline
\end{tabular}
\end{table*}

\subsection{MERLIN observations of the neutral hydrogen absorption}

The 512$\times$512$\times$63 spectral line cube was derived using
natural weighting and has a resolution of 0.27
$\times$ 0.18 arcsec$^2$ and an RMS noise, $\sigma$, of 0.51 mJy
beam$^{-1}$ over absorption-free channels of the cube. As shown in
Figures 3 and 4, neutral hydrogen absorption is only detected towards the
south-western (SW) radio emission. From Gaussian fitting to the absorption
spectra we measured the centroid velocity (optical heliocentric
convention) and half power width (FWHM) of the absorbing gas. The opacity
and column densities were obtained directly from the spectra (we assumed a
spin temperature of 100K in calculating the column density).

\begin{figure*}
\centering
\psfig{figure=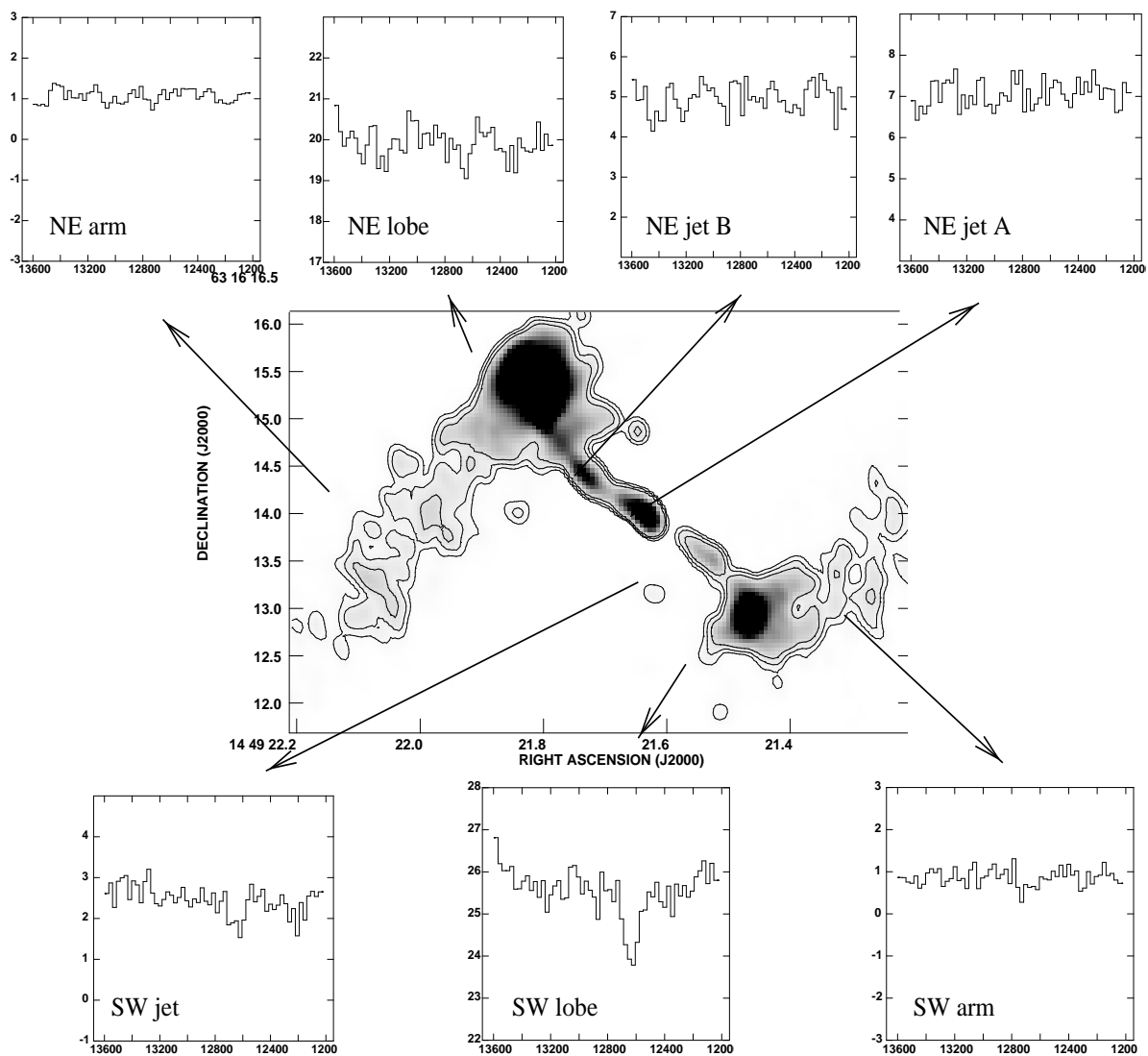,width=16cm,height=15cm}
\caption{A montage showing the 1.4 GHz continuum image of 3C\,305 and spectra 
towards the continuum components. The grey scale is from 
0 to 10 mJy beam$^{-1}$ and the contours levels are (0.3,0.6,1.2) mJy
beam$^{-1}$, the angular resolution is $0.19 \times 0.14$ arcsec. The
spectra are plotted with flux density in mJy beam$^{-1}$ against
optical-heliocentric velocity in km s$^{-1}$. Neutral hydrogen is 
detected towards only the southwest jet and lobe.}
\label{fig2}
\end{figure*}

The non-detection of absorption towards the northeastern jet components
A and B puts constraints on the column densities of \hone\ in front of
these features.  Limits for the opacity and column densities were
calculated from the RMS noise in the
spectra and an expected line width of 150 km s$^{-1}$. Spectra in the
direction of the bright northeast lobe are limited in dynamic range at 
the 1\% level due to baseline uncertainties. The absorption results are 
summarised in Table 4.

\begin{figure*} \centering \psfig{figure=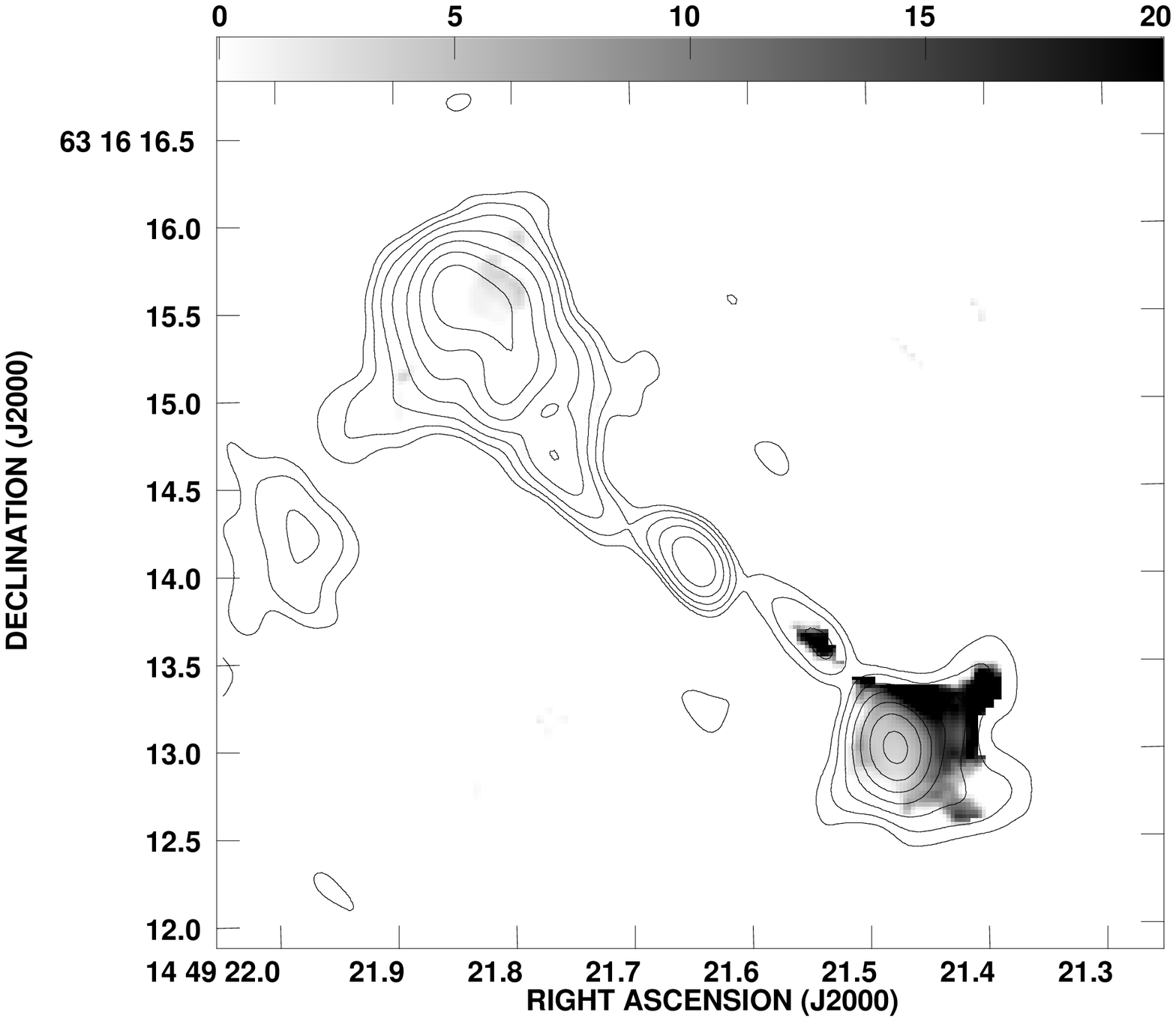,width=10cm}
\caption{The integrated neutral hydrogen absorption optical depth
superimposed on a contour map of the 20cm continuum. The contour levels
are 1,2,4,8,16,32,64) mJy beam$^{-1}$. The greyscales shows the integrated
\hone\ optical depth ranging from 0 to 20 km~s$^{-1}$. Both sets of
measurements were derived from the naturally weighted data.} \label{fig3}
\end{figure*}

%\begin{figure*}
%\centering
%\psfig{figure=MOM1,width=10cm}
%\caption{The average (moment 1) velocity of the neutral hydrogen
%absorption over the
%SW lobe. The grey scale is from 12620 km s$^{-1}$ (white) to 12740 km
%s$^{-1}$ (black) and the contours are in intervals of 20 km s$^{-1}$.}
%\label{fig4}
%\end{figure*}

\begin{table*}
\centering
\caption{The properties of the neutral hydrogen absorption in the central region of 3C\,305.}
\label{tab1}
\begin{tabular}{lcccccc}
\hline
\hline
Component    & V$_{\rm{c}}$  & $\delta$V        & $\int{\tau}dV$ & N$_{\rm{H}}$ $\times$ 10$^{21}$ \\
             & (km s$^{-1})$ & (km s$^{-1}$)    &(km s$^{-1})$   & ( atom cm$^{-2}$) \\
\hline

Ne Arm       &               &                  & $<$12.3        & $<$2.2            \\
NE Lobe      &               &                  & $<$0.7         & $<$0.1            \\
NE Jet B     &               &                  & $<$6.0         & $<$1.1\\
NE Jet A     &               &                  & $<$2.1         & $<$0.4\\
SW Jet       &12660$\pm$26   & 130$\pm$26       & 70$\pm$24      & 12.8$\pm$4.4      \\
SW Lobe      &12640$\pm$26   & 145$\pm$26       & 11$\pm$4       & 1.9$\pm$0.7       \\
SW Arm       &               &                  & $<$19.9        & $<$3.6            \\
             &               &                  &                &                   \\
\hline
\end{tabular}
\end{table*}

\subsection{HST observations}

\begin{figure*}
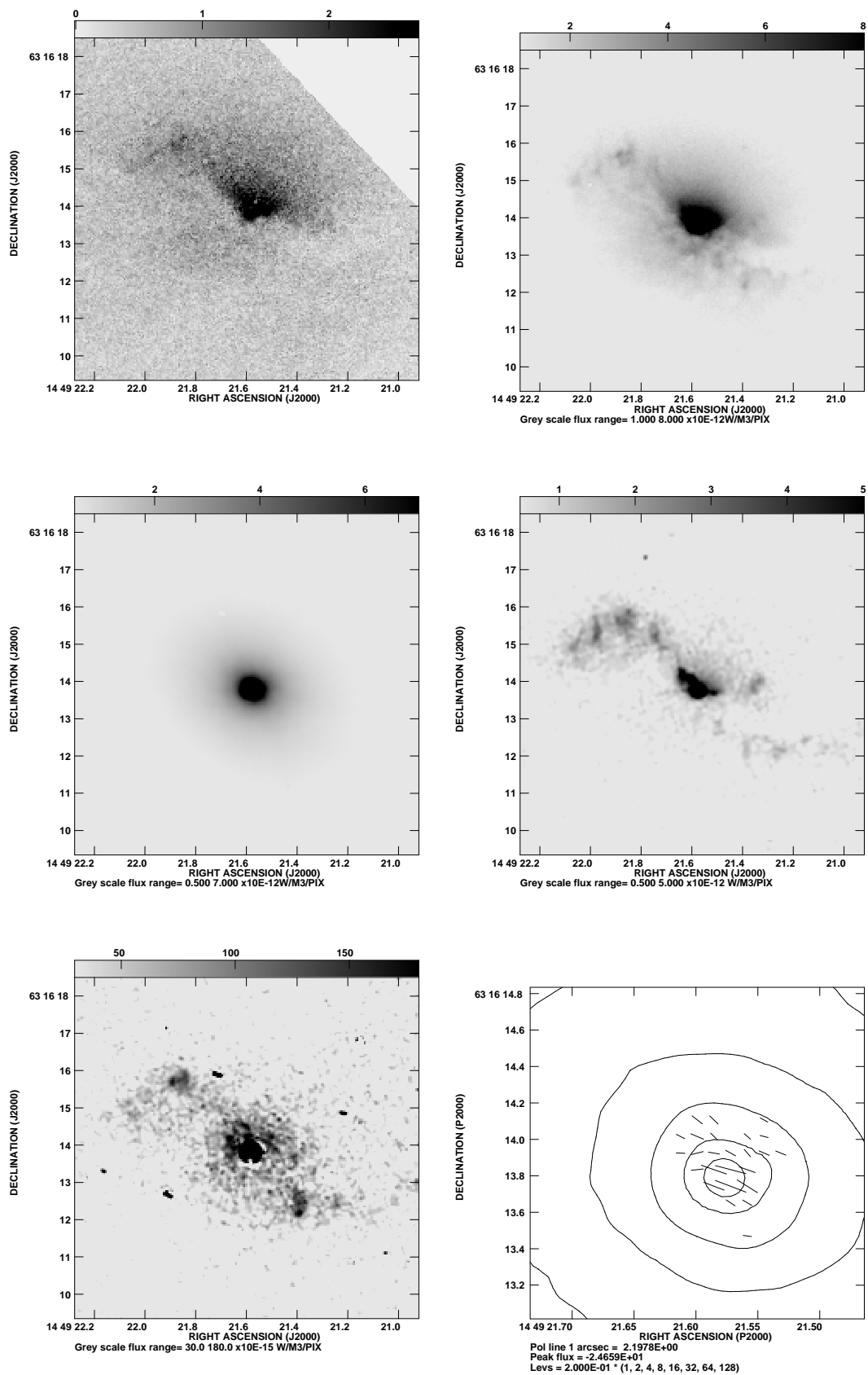

\begin{tabular}{cc}
\psfig{figure=FIG430,width=6.85cm}&
\psfig{figure=FIG702,width=6.85cm}\\    
\psfig{figure=FIG160,width=6.85cm}&     
\psfig{figure=FIGO3,width=6.85cm}\\
\psfig{figure=FIGFEII,width=6.85cm}&
\psfig{figure=FIGPOL,width=6.85cm}\\
\end{tabular}
\caption{HST images of 3C\,305. Top left: FOC 430-nm image. Top right:
WFPC2 F702W image, containing H$\alpha$ and the surrounding continuum.
Middle left: NICMOS F160W $H$-band image. Middle right: narrow-band image
around the \othree\ line. Bottom left: narrow-band image of infra-red
\fetwo 1.644$\mu$m. Bottom right: zoomed-in view of polarization   
vectors at $K$-band in the nuclear region. This is the only part of the
source in which significant polarization is visible.}
\end{figure*}  

In Fig. 5 we show all the optical images of 3C\,305 as greyscale plots. 
Immediately obvious from these images is the contrast between the 
extreme smoothness of the infra-red images, dominated by the old galaxy
component, and the irregular structures seen in the blue continuum  
and line images. Emission in the images of the \fetwo\ and \othree\ lines
extends along the radio axis. At the end of the northeastern
radio jet, emission in both lines is seen. In the case of \othree\ the
emission extends some distance beyond the head of the jet in a
southeasterly direction. The \othree\ emission is fainter on the
counterjet (southwest) side, but turns northward after the end of the 
radio jet; the overall effect is of a 'Z'shape in the \othree\ emission.
The \fetwo emission is relatively much brighter at the ends of the
jet, and fainter in the regions away from the main jet axis.

Faint extended emission is detectable in the $K$-band images which were
originally used to collect polarization information. We interpret this
as Pa$\alpha$ at rest wavelength 1.879$\mu$m, as it is not seen in
the continuum $H$-band image. A smoothed version of the $K$-band image has
been subtracted to reveal the underlying line emission (Fig. 6).

\begin{figure}
\psfig{figure=FIGH2,width=8cm}
\caption{Pa$\alpha$ emission obtained by subtracting a convolved
and scaled version of the $K$-band data from the $H$-band data.}
\end{figure}

The dust lane on the southern side (Heckman et al. 1982, Jackson et al.
1995) is clearly visible in the colour images, and is particularly
prominent and sharp-edged in the 430-nm image. Patchy dust obscuration
is obviously present throughout the central and southern regions of the
nuclear source.

The polarization images (bottom right of Fig. 5, and Fig. 7) 
show very similar structure to that seen by   
Draper et al. (1993). The peak polarization is about 2\% and is   
concentrated to the north of the optical core, on the side unaffected by
the dust lane.

\begin{figure}
\psfig{figure=FIGPPOL,width=6.85cm}
\caption{Percentage polarization of the core of 3C\,305. The polarization
level agrees roughly with that found by Draper et al. (1993) a little
further from the centre, both in magnitude and direction. The grey-scale
runs from 0 to 5\% and is blanked where no significant detection is
made. Contours correspond to steps of factors of 2 in the total-flux image.}
\end{figure}

Unsharp masking was applied to the F702W image 
by subtraction of a 0.5 arcsec Gaussian
smoothed image at 50\% intensity. This process removes low spatial frequency
variations in the image and enables more detail to be seen. The unsharp
masked image is shown in Fig. 8 and reveals a complex nuclear dust
lane.

\begin{figure*}
\centering
\psfig{figure=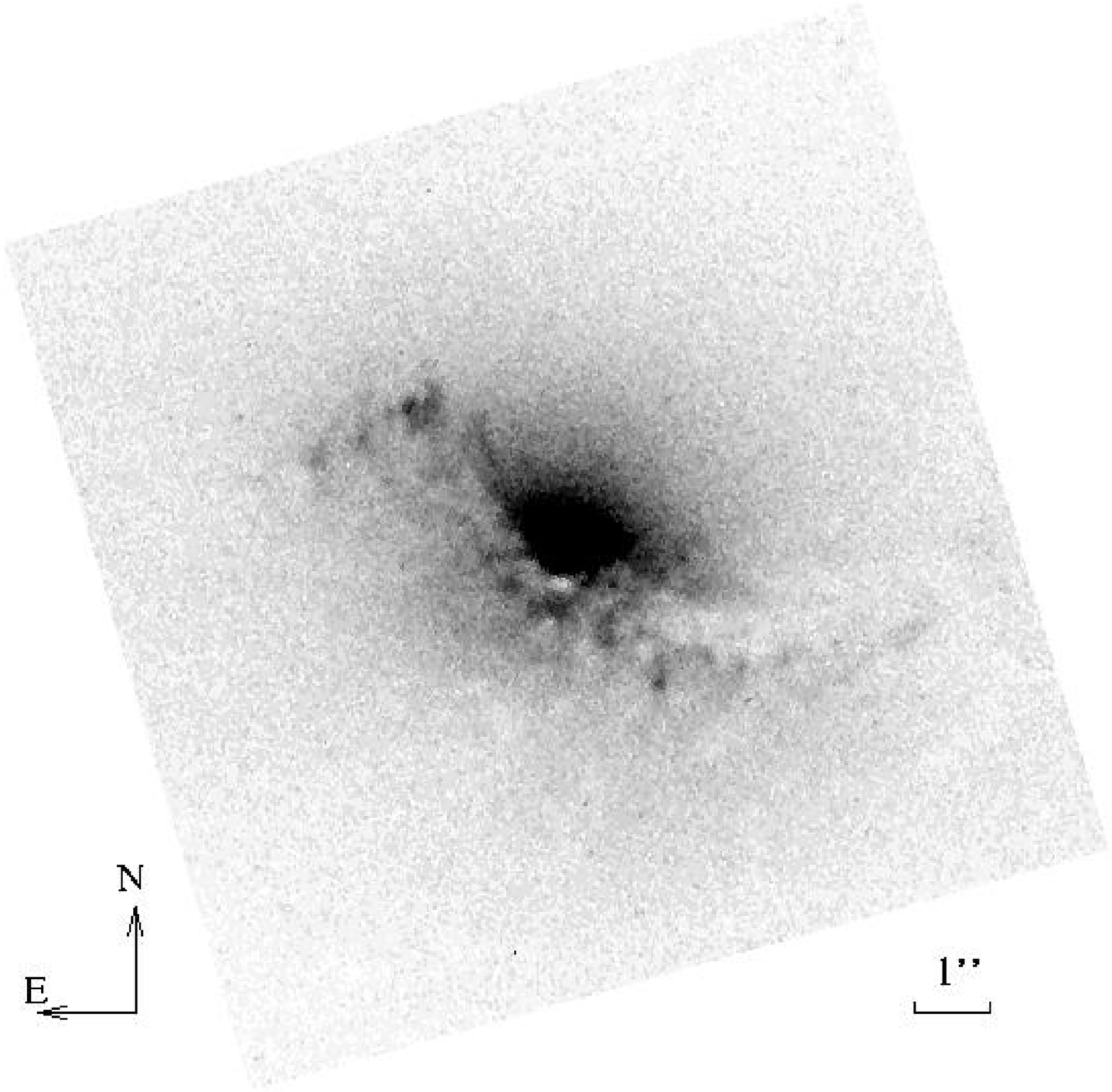,width=16cm}
\caption{HST unsharp masked image of the central region of 3C\,305 revealing the peculiar shape of the nuclear dust lane.}
\label{fig5}
\end{figure*}

\section{Discussion}

\subsection{Detection of an unresolved nucleus}

In the optical part of the spectrum, the central galaxy profile appears
smooth. Chiaberge, Capetti \& Celotti (1999) attempted to detect a
compact optical core in the 702-nm HST image, by fitting a smooth
profile to regions $>$0\farcs23 from the centre, extrapolating the fit
to the central regions and testing for any remaining central component of
width $\leq$0\farcs08. They found none, as did Hardcastle \& Worrall (2000)
using the same image. We have extracted the 702-nm image from the HST
archive and subtracted TinyTim point-spread functions (Krist 1995) from
it until an unphysical ``hole'' is clearly detected in the intensity
distribution. This results in a limit of 
$<3.5\times10^{-19}$W m$^{-2}$nm$^{-1}$ ($m_{R}>19.5$) for the optical 
$R$-band magnitude of any central unresolved object.

Chiaberge et al., however, did find optical point sources in a number of
other low power, Fanaroff \& Riley type 1 (FR~I, Fanaroff \& Riley 1974)
radio galaxies. In other low-power radio galaxies such as
Centaurus A (Schreier et al. 1998) as well as higher-power FR~II galaxies
such as Cygnus A (Tadhunter et al. 2000) nuclear infra-red point sources
have now been detected at HST resolution.

In both the $H$ and $K$-band NICMOS images we see evidence for an
unresolved nuclear source on scales of $<$50\,pc. Both images have been
fitted by a cusped central galaxy profile, a dependence of brightness
with radius $I(r)$ which has been successfully
applied by Faber et al. (1997) to HST images of a wide range of
early-type galaxies,

\[
I(r) = I_{b}2^{(\beta-\gamma)/\alpha}(r_b/r)^{\gamma}(1-(r/r_b)^{\alpha})^{(\gamma-\beta)/\alpha},
\]

where $I_b,r_b,\alpha,\beta$ and $\gamma$ are constants. We have ignored
the central 0\farcs5 of the images when performing this fit. A downhill 
simplex fit has been used for simultaneous fitting of all parameters 
(e.g. Press et al. 1992) and the results are given in Table 5. 
A central condensation, in addition to the galaxy fit, is
then seen in both images which has the characteristic form of the NICMOS
point spread function (Fig. 9). We have normalised the photometry
to the large-aperture results of Impey (1983) who measured the $H$-band
and $K$-band flux densities through large (8\farcs2 and 12\farcs3) apertures.
This central image has a $K$-magnitude of about 16.3$\pm$1.0  and an
$H$ magnitude of 14.7$\pm$0.3. Extrapolating from the $K$-band result using
a typical quasar colour of $V-K\sim 3$ gives $M_V\sim -22.5$. This is
typical of luminosities required to photoionize extended nebulosities in
nearby objects, although we argue in section 4.5 that jet-driven shocks
may also play a major role in exciting the gas.

\begin{table}
\begin{tabular}{lcc}
Parameter & $H$-band & $K$-band\\
&&\\
Ellipticity & 0.85 & 0.81 \\
Position angle of major axis & 67 & 66 \\
Break radius, $r_b$ (mas) & 470 & 190 \\
$\alpha$ & 0.44 & 1.96 \\
$\beta$ & 1.68 & 1.23 \\
$\gamma$ & 0.53 & 0.30\\
\end{tabular}
\caption{Best-fit parameters from a cusp-type model (see text) 
for the fitting of the 3C\,305 host galaxy
from the HST images in the $H$-band (F160W) and $K$-band (POL) filters.}
\end{table}

\begin{figure}
\psfig{figure=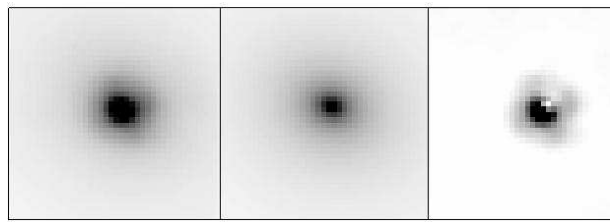,width=8cm}
\caption{From left to right: the NICMOS $H$-band data, the best-fit
model from the cusp-law model used by Faber et al. (1997) fitted to the
data outside the central region, and the residual. The latter clearly 
shows the NICMOS point-spread function.}
\end{figure}

The natural inference from these observations is that we are indeed
detecting the central AGN in 3C\,305 in the infrared. A similar conclusion
has been reached in ground-based images of a sample of 3CR radio
galaxies by Simpson, Ward \& Wall (2000) who detected excess flux in 5
out of 10 sets of infrared images. Simpson et al.'s observations were
ground-based, and their analysis relied on separation of the central
nuclear light by fitting a de Vaucouleurs profile together with an
unresolved component. Here we are able to resolve down to 0\farcs07 and
determine the parameters of the nucleus directly.

The infrared colour of the nucleus of 3C\,305 is roughly comparable 
to the typical $J-K\sim 2.5$ found by Simpson et al. However, with the 
$R$-band limit we can derive an estimate for the reddening. We assume
that the hidden nucleus is a quasar, and furthermore assume a typical
optical-IR spectral energy distribution 
represented by a power-law, $F_{\nu}\propto\nu^{-\alpha}$.
Using the $H$- and $R$-band results, we find for $\alpha$=0.5 that the
central reddening $A_{V}>4$ magnitudes. This is a relatively modest
limit beside those derived from infrared emission-line (Ward et al. 
1991) and X-ray (Ueno et al. 1994) observations of objects such as
Cygnus A. However, the $H$-band continuum flux may be coming from a 
relatively large area and be subject to less obscuration than the 
central part of the nuclear region. Further deep observations in the 
red optical and near-IR would establish the point-source spectral 
energy distribution and hence the obscuration to the nuclear component 
more definitively.

The astrometry we carried out (section 2.3) shows that the brightest part
of the radio image is not coincident with the optical/infrared nucleus
discussed above. Hence this radio component is most likely a knot in the
jet, certainly not a radio nucleus, and hence we shall refer to this as NE
jet A. The optical nucleus (Figure 2), appears to be situated in the gap
between this component and the SE jet and hence has a radio flux density less
than 1 mJy. This could be a consequence of either synchrotron self
absorption, or free-free absorption by nuclear ionised gas with an
emission measure of $\sim 10^7$ pc cm$^{-6}$. It could also be a
consequence of relativistic beaming; if the radio plasma is outflowing
with a Lorentz $\gamma$ of about 5, then Doppler deboosting of intrinsic
flux takes place if the outflow is further than about 60$^{\circ}$ to
the line of sight. Such an angle would still maintain the observed 
factor of a few in relative boosting of the flux of the northeastern 
jet relative to the southwestern jet, if the NE jet was the approaching 
radio jet.

Clearly \hone\ absorption cannot be measured in the direction of the
nucleus, and hence no limits can be set to atomic
hydrogen column densities associated with a dusty torus on scales of
$\sim$1--10~pc.

\subsection{The structure of the neutral gas and dust}

We can use the HST observations to determine the amount of dust
reddening present and the MERLIN \hone\ absorption observations to determine
the neutral gas column in front of the radio source. If these are compatible,
assuming standard dust-to-gas ratios, it means that the dust as well as the
gas must lie in front of the radio source; hence we can make deductions
about the overall geometry of the radio source. We consider both
components in turn.

\subsubsection{Dust}
A distinct dust lane extends over the central region of 3C\,305, offset
0\farcs4 to the SE of the nucleus (see Fig. 8). The dust lane 
splits into two distinct structures, about 0\farcs6 wide, extending 
1\farcs5 to the northeast of the nucleus. The dust lane is most 
prominent to the southwest of the nucleus and extends 4\farcs5 to the 
southwest of the nucleus before bending sharply.

The apparent cone-like shape of the optical nucleus perpendicular to the
radio axis is explained by the obscuration of the emission in the
southwest of the source by the dust lane. Dust obscuration is probably also
responsible for the apparent highly localised ($<$0.25 arcsec$^2$) regions
of emission to the S and SW of the nucleus. Over the central 2 arcsec, the
structure appears to be consistent with an inclined disc of dust encircling
the active galactic nucleus at an inclination of $\sim$ 45$^{\circ}$.

\begin{figure}
\psfig{figure=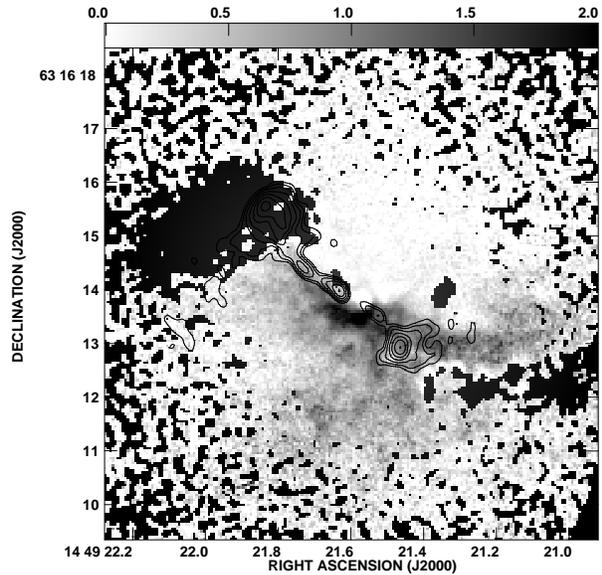,width=8cm}
\caption{Reddening map of 3C\,305, derived from the 1.6-$\mu$m and 702-nm
images as described in the text.
Grey-scale represents values of $A_{V}$ from 0 to 2 magnitudes and
assumes that all obscuration is in front of the emission. Areas of
strong optical line emission, which contaminate the 702-nm image, have
been blanked and appear black.}
\end{figure}

In order to quantify the amount of dust present, we have generated a map
of extinction $A(V)$ using the 1.6-$\mu$m image together with the 702-nm
image. When these images are divided, the northern segment shows smooth
undisturbed contours, and these have been assumed to represent the
unobscured emission of the galaxy. This has been fitted with an
elliptically symmetric function, divided out, and the extinction map
in magnitudes has been formed by taking the logarithm and multiplying by
2.5. Assuming that the extinction $A_{1.6\mu{\rm m}}-A_{702{\rm nm}}=-1.66$E(B$-$V), a
reddening map of $A_V$ can be derived if we assume Galactic properties for
the dust; the result is shown in Fig. 10. The typical $A_V$ is about 1.5
magnitudes across much of the source, implying a gas column of
about 3.4$\times10^{21}\,$cm$^{-2}$ assuming Galactic gas-to-dust
ratios. In particular, this is the predicted gas column against the SW
radio lobe.

It should be noted, however, that the derived E(B$-$V) for the dust is a
lower limit. If the dust column were not all situated in front of the
majority of the galaxy emission, much larger values of E(B$-$V), and
hence larger gas columns, would be
needed. The comparison of the dust prediction with direct HI
measurements can therefore be used directly to constrain the geometry in
this case.

\subsubsection{Gas}

We have detected neutral hydrogen absorption against the southwestern jet
and lobe which confirms the presence of neutral gas in front of these radio
structures (see Fig. 3). The absorption is localised to the southwest side
only, which we attribute to gas associated with a peculiar shaped dust lane
(see Fig. 10). The fact that absorption is seen against the southwestern
jet agrees with the plausible assumption that the northeastern radio jet
has a component of its velocity towards us and appears brighter than the
SW jet due to Doppler boosting.

The average column density of the neutral hydrogen against the southwestern
lobe is high, N$_{\rm{H}}$ = (1.9 $\pm$0.7)$\times$10$^{21}$ cm$^{-2}$ and is
similar to column densities found against the central regions of the
radio galaxy 3C\,293( $\sim$ 1.7 $\times$10$^{21}$ cm$^{-2}$ - Beswick et
al 2002). This agrees within a factor of 2 with the column density
derived above (section 4.2.1) using optical reddening. Given the
assumptions of both methods, such as a spin temperature of 100K in the
radio method and a galactic dust to gas ratio in the optical method, there
is reasonable agreement in the column density estimates. This is
consistent with the atomic hydrogen being associated with optical dust
lane and that the dust lane is in front of the southwestern radio lobe.
The dust lane extends over an area of $\sim$3.2 arcsec$^2$ out to 4.5
arcsec southwest of the nucleus. Assuming an average column density over this
region of 1.9 $\times$10$^{21}$ cm$^{-2}$, the total mass of neutral
hydrogen in the dust lane is 3 $\times$10$^{7}$ M$_{\odot}$.

Our results help to explain the observations of Draper et al. (1993)
and Jackson et al. (1995). Draper et al. investigated the polarization
structure of 3C\,305 and concluded that the lack of coincidence between
the polarised regions and the emission line regions or radio jet axis
was due to obscuration within the host galaxy. Jackson et al. found
the southern side of the nucleus was slightly redder in colour which
they also interpreted as a consequence of obscuration.

\subsection{Kinematics of the Neutral Gas and Dust}

The centroid velocity of the absorption line (12640$\pm$26 km s$^{-1}$)
is red-shifted by 130 km s$^{-1}$ relative to the systemic velocity of
(12510$\pm$60)  km s$^{-1}$ (Heckman et al. 1982). Optical spectroscopy by
Heckman et al. (1982) obtained velocities for the \othree\ emission along PA
57$^{\circ}$ with a velocity resolution of $\sim$130 km s$^{-1}$.

There is a clear agreement between the emission-line velocity and the
neutral hydrogen velocity implying that the kinematics of the two
structures are linked.  The neutral hydrogen absorption is localised on
scales of $\sim$2 arcsec as can be seen in Fig. 3. The velocities of the
absorbing gas are shown in Fig 4. The velocity of the absorbing components
varies over the region by $\sim$ 75 km s$^{-1}$ arcsec$^{-1}$. From Fig.
10 of Heckman et al. (1982), the velocity gradient of the emission line
gas along PA 57$^{\circ}$ over the central 6.5 arcsec is $\sim$80 km
s$^{-1}$ arcsec$^{-1}$.

The kinematics of the neutral and ionised gas are inconsistent with the
stellar dynamics which represent only $\sim$55\% of the rotation velocity
of the ionised gas as found 
by Heckman et al. (1985). This is consistent with
Heckman's original (1982) suggestion that the neutral gas is disturbed by
the radio jets.

%BROAD ABSORPTION LINES?

\subsection{Geometry of 3C\,305}

From our neutral hydrogen absorption observations we can infer that the SW
radio structure lies behind the nuclear dust lane. However no absorption
is detected against the stronger northeast lobe despite the presence of dust
lanes. We interpet this as evidence that the northeast lobe is in front of the
neutral gas. Thus we suggest that the radio emission is towards the
observer in the northeast and away from the observer in the southwest.

The kinematics of the neutral gas and dust appear to be closely
linked with the dynamics of the ionised gas (Heckman et al. 1982). Considering
the apparent disc-like structure of the dust lane in the central few
arcsec, we suggest that both the neutral and ionised gas are undergoing
circular motion around the centre of the galaxy.

\subsection{Emission lines and interactions of the radio jet}

The mechanism by which emission lines in active galaxies are excited is
complex and not well understood. A possible source of ionization is the
emission of hard-UV and X-ray photons from the central AGN. Such a nucleus
may well be present in 3C\,305 although as we have argued above the 
obscuration may prevent us from seeing the nucleus directly.

The other possibility is that shocks induced by the radio jet's passage
through the intergalactic medium photoionize the gas (Sutherland, Bicknell \& Dopita
1993). The primary physical mechanism in this case is still
photoionization, but in this case carried out by the high-energy photons
produced in the shock. Because of the relative similarity of the two
processes, it is not easy to distinguish which is actually responsible
for the observed line emission. One diagnostic may be the infra-red
\fetwo line, which is known to be produced abundantly in fast shocks
associated, for example, with supernova remnants (Moorwood \& Oliva
1988). Forbes \& Ward (1993) suggested that shocks may also be
responsible for \fetwo emission, the strength of which is correlated with
radio core flux densities in Seyfert galaxies.

Simpson et al. (1996) have investigated the production of \fetwo in
detail and present diagnostic diagrams involving the ratios of 
\fetwo/Pa$\beta$ and \oone/H$\alpha$. They find more scatter between
\fetwo and radio fluxes than Forbes \& Ward and suggest in particular
that normal AGN photoionization can account for the observed ratios of
\fetwo/Pa$\beta$ in their sample.

In 3C\,305 we have an opportunity to study this in detail. The reason is
that in this object there is a close association of the northeastern
termination of a radio jet with a strong region of line emission. Such a
structure would be expected to produce line ratios corresponding to
shock excitation.

In Fig. 11 we show a slice through the northeast section of the
emission line region. It is apparent that the profiles of \fetwo and
\othree are different; there is a clear peak of \fetwo around the point
where the jet terminates, and \othree is more evenly distributed
throughout the emission line region. This strongly suggests that we are
here seeing an increase in \fetwo flux due to shock ionization around an
interaction region.

The H$\alpha$/\fetwo flux ratio in this region is about 17. Assuming
case B values for the hydrogen line emission, and converting from our
\fetwo $\lambda 1.644\mu$m fluxes to the 1.257-$\mu$m line used by
Simpson et al., we obtain a ratio \fetwo$_{1.257\mu{\rm m}}$/Pa$\beta$
of just over 1. Adopting Heckman et al.'s (1982) value of $\sim$0.13 for
the radio \oone/H$\alpha$ puts the strong emission-line region in 3C\,305
approximately in the region of Simpson et al.'s diagnostic diagram where
standard Seyfert emission lines lie. This implies in turn that standard
Seyferts may have some contribution to their line emission from
jet-induced shocks, although this argument does not of course prove that
this is the case.

\begin{figure}
\psfig{figure=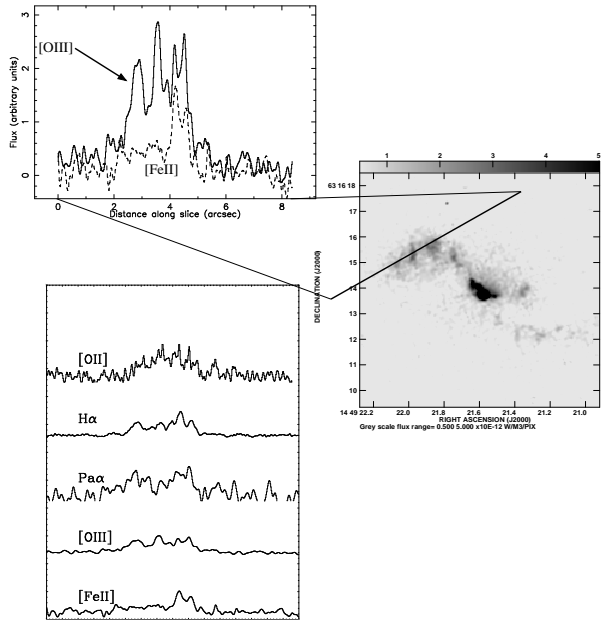,width=8cm}
\caption{Slice through the northeast knot complex, in \othree\ (solid
line) and infra-red [\fetwo] (dashed line). The bottom panel includes all
of the emission lines studied. 
The profiles of H$\alpha$ and \otwo\ have had a continuum subtracted by
eye, and it has therefore been assumed that only emission lines
contribute to the compact condensations in the northeastern lobe.}
\end{figure}

\section{Conclusion}

The major results of this paper are the discovery of a pointlike
infrared nucleus; the identification of the nuclear position which is
not coincident with the bright central radio component; neutral hydrogen
gas rotation velocities of 130kms$^{-1}$ relative to systemic velocity, 
aligned with, and sharing a common velocity gradient with, the emission 
line gas; the comparison of dust and gas columns which shows that 
most of the dust associated with the emission line gas is in front of 
the southwestern part of the radio source (and hence that the southwestern
radio jet is almost certainly the receding jet); and the discovery of
peaks of \fetwo\ 1.64$\mu$m emission coincident with radio hotspots,
consistent with the presence of jet-induced shocks.

There is no evidence from the new infrared picture that 3C\,305 contains
a double nucleus as in the case of the merging system Mkn 463 (Hutchings \&
Neff 1989), and the infrared isophotes are quite smooth indicating a
relatively undisturbed old stellar population, such as might be expected
$>10^8$ yr from the start of a merger process. The point infrared
nucleus is consistent with a central QSO which may have formed during
the course of the merger.

\section{Acknowledgements}

We thank Simon Garrington and Tom Muxlow for assistance with the
observations and Ian Browne for useful comments. 
We are very grateful to R. Argyle and L. Morrison for their work in fixing
the astrometry of the 3C\,305 field.

This research was based on observations with the Hubble Space Telescope,
obtained at the Space Telescope Science Institute, which is operated by
Associated Universities for Research in Astronomy, Inc., under NASA
contract NAS5-26555; and with MERLIN, which 
is operated as a National Facility by the
University of Manchester on behalf of the UK Particle Physics \&
Astronomy Research Council.

{}

\end{document}